\documentclass[12pt]{article}
\usepackage{times}

\usepackage[labelfont=bf]{caption}
\usepackage[superscript,nomove]{cite}

\usepackage[margin=1in]{geometry}

\usepackage{graphicx}
\usepackage{lineno}
\usepackage{nameref}

\def\msol{\;\mathrm{M_\odot}}
\def\kpc{\;\mathrm{kpc}}
\newcommand{\unit}[1]{\;\mathrm{#1}}

\makeatletter
\renewcommand{\maketitle}{\bgroup\setlength{\parindent}{0pt}
	\begin{flushleft}
		\textsf{\textbf{\Large{\@title}}}

		\@author
		
		\emph{\@date}
	\end{flushleft}\egroup
}
\makeatother

\makeatletter
\renewcommand\@biblabel[1]{#1.}
\makeatother

\title{The Magellanic Corona and the\\ formation of the Magellanic Stream}

\author{S. Lucchini,$^{1}$ E. D'Onghia*,$^{1,2,3}$ A. J. Fox,$^{4}$ C. Bustard,$^{1}$ J. Bland-Hawthorn$^{5,6}$ \& E. Zweibel$^{1,2}$}
\date{
	$^1$ Department of Physics, University of Wisconsin-Madison, 1150 University Avenue, Madison, WI, USA 53706.\\
	$^2$ Department of Astronomy, University of Wisconsin-Madison, 1150 University Avenue, Madison, WI, USA 53706.\\
	$^3$ Center for Computational Astrophysics, Flatiron Institute, 162 Fifth Avenue, New York, NY, USA 10010.\\
	$^4$ AURA for ESA, Space Telescope Science Institute, 3700 San Martin Drive, Baltimore, MD, USA 21218.\\
	$^5$ Sydney Institute for Astronomy, School of Physics, University of Sydney, NSW 2006, Australia.\\
	$^6$ ARC Centre of Excellence for All Sky Astrophysics in 3D, Australia.
}

\begin{document}

\maketitle



{\bf \noindent
The dominant gas structure in the Galactic halo is the Magellanic Stream, an extended network of neutral and ionized filaments surrounding the Large and Small Magellanic Clouds (LMC/SMC), the two most massive satellite galaxies of the Milky Way\cite{Mathewson1974,Nidever2008}. Recent observations indicate that the Clouds are on their first passage around our Galaxy\cite{Kallivayalil2013}, the Stream is made up of gas stripped from both the LMC and the SMC\cite{Nidever2008,Fox2013,Richter2013}, and the majority of this gas is ionized\cite{Fox2014,Barger2017}. While it has long been suspected that tidal forces\cite{Besla2012,Pardy2018} and ram-pressure stripping\cite{Hammer2015,Wang2019} contributed to the Stream's formation, a full understanding of its origins has defied modelers for decades\cite{D'Onghia2016}. The recently-determined high mass of the LMC\cite{Penarrubia2016} and the detection of highly ionized gas toward stars in the LMC\cite{Wakker1998,Lehner2009} suggest the existence of a halo of warm ionized gas around the LMC. Here we show that by including this ``Magellanic Corona" in our hydrodynamic simulations of the Magellanic Clouds falling onto the Galaxy, we can simultaneously reproduce the Stream and its Leading Arm. Our simulations explain the Stream’s filamentary structure, spatial extent, radial velocity gradient, and total ionized gas mass. We predict that the Magellanic Corona will be unambiguously observable via high-ionization absorption lines in the ultraviolet spectra of background quasars lying near the LMC. This prediction is directly testable with the Cosmic Origins Spectrograph on the Hubble Space Telescope.
}



\begin{figure}
	\centering
	    \includegraphics[width=\textwidth]{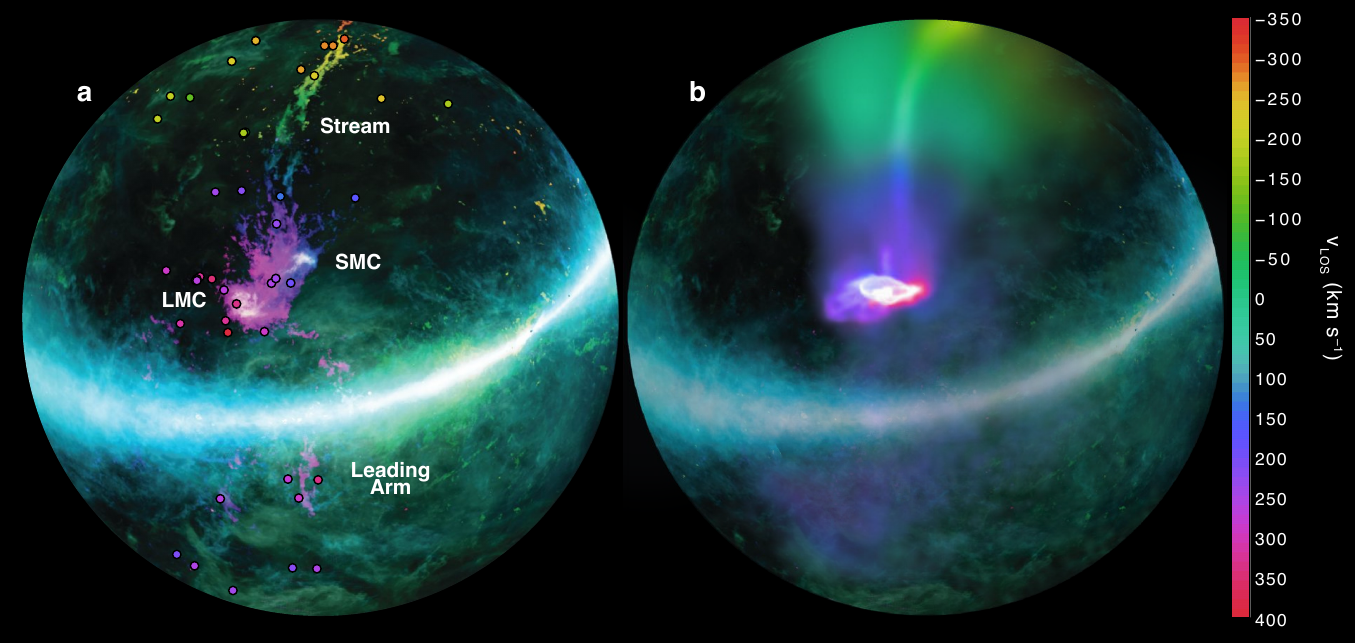}
	\caption{{\bf The Magellanic Stream in zenithal equal-area coordinates}.
	{\bf a}, Observed H\,I data\cite{McClure-Griffiths2009} of the Magellanic Stream with line-of-sight velocity displayed by the color bar (from $-$350 km\,s$^{-1}$ to 400 km\,s$^{-1}$) and brightness indicating the relative gas column density. The points represent the sightlines with UV-absorption-line observations from the \emph{Hubble Space Telescope}\cite{Fox2014} colored by their line-of-sight velocity. These points show the extent of the ionized gas associated with the Stream. {\bf b}, The results of the model including the Magellanic Corona and the MW hot corona. Gas originating in both the LMC and SMC disks is shown in the model without separating the neutral gas from ionized gas. This affects the morphology of the Stream, causing the model to appear smoother and less fragmented than the data. However, the model reproduces the current spatial location and velocity of both Clouds, and the velocity gradient of the gas along the Stream. The Milky Way disk and background are extracted from real H\,I images\cite{McClure-Griffiths2009}.
	}
	\label{fig:zea}
\end{figure}

Our Galaxy is accompanied by two fairly massive dwarf galaxies, the Large and Small Magellanic Clouds, and a massive gaseous tail trailing behind them, the Magellanic Stream. The Stream is an interwoven tail of filaments pulled out of the Magellanic Clouds (MCs) in their orbit around the Milky Way (MW)\cite{Mathewson1974,Bruns2005,Nidever2008}.
When considered together with its Leading Arm (LA) -- the Stream's counterpart in front of the MCs -- the Stream stretches over 200 degrees on the sky (see Fig. \ref{fig:zea}a)\cite{Nidever2010}.
With a total mass of $\sim1-2\times10^{9}\msol$ (consisting of $\sim2\times10^{8}\msol$ neutral hydrogen and the remainder in ionized gas)\cite{Fox2014, Barger2017} the Magellanic Stream dominates all the other gas clouds in the Galactic halo, both in terms of gas mass and gas inflow rate. Therefore understanding the Stream is essential to a global picture of the Galaxy's circumgalactic medium\cite{D'Onghia2016}.

The current paradigm of Stream formation is known as the first-infall model\cite{Besla2007, Besla2012}. In this scenario, tidal forces from the LMC acting on the SMC when the Clouds are at their first pericentric passage around the MW lead to the formation of the Stream. This model is motivated by the high tangential velocities of the Clouds\cite{Kallivayalil2013} and the strong morphological disturbances observed in the SMC\cite{Bruns2005,Besla2012,Pardy2018}.
This model successfully reproduces the size and shape of the Stream, but several difficulties remain\cite{D'Onghia2016}:
{\it (i)} the observed Stream is significantly more extended spatially and a factor of up to ten more massive than the simulated Stream, especially when including its ionized component, which dominates the mass budget\cite{Fox2014}; 
{\it (ii)} the fragmented structure of the Stream and Leading Arm indicates that the interaction with the MW gas corona plays a significant role and cannot be ignored;
{\it (iii)} the Stream is bifurcated, with kinematic and chemical analyses indicating that gas from both the LMC and SMC is present\cite{Nidever2008,Fox2013,Richter2013}. This indicates that the Stream has a dual origin, whereas tidal models predict an SMC origin because of the shallower potential well of the SMC.

The discovery of several ultra-faint dwarfs around the LMC\cite{Bechtol2015} indicates that the LMC and SMC likely entered the MW recently as part of a system of dwarf galaxies (the Magellanic Group), with the LMC as its largest member\cite{D'Onghia2008,Nichols2011}.
Given the LMC dark matter halo mass of $\sim2\times10^{11}\msol$ \cite{Penarrubia2016}, the virial temperature is $\sim5\times10^{5}\unit{K}$, so the Magellanic Group is expected to contain a warm gas corona at this temperature. Furthermore, cosmological simulations of MW-sized galaxies with LMC-like dwarf satellites \cite{Pardy2019} predict the existence of ionized gas halos surrounding those satellites\cite{Hafen2019}. The presence of an LMC corona is also motivated by detections of absorption from highly-ionized carbon in ``down-the-barrel'' spectroscopic observations of hot stars in the LMC \cite{Wakker1998,Lehner2009}.
Such coronae are likely kept warm via energy input from stellar feedback and outflows. 

Here we show that by including this ``Magellanic Corona'' in hydrodynamic simulations of Stream 
formation, the mass budget discrepancy of the Stream can be solved. Crucially, we reproduce the 
ionized component for the first time. The Magellanic Corona appears to be the key missing ingredient in models of Stream formation.
The new simulations shown here have been run with GIZMO Hydrodynamic N-Body code including radiative cooling and heating, star formation, and stellar feedback to model the LMC-SMC-MW dynamics including the Magellanic Corona (see Methods). During the initial stages of the LMC-SMC tidal interaction, the pair lie outside the MW's gravitational influence. The cold gas in the extended disk of the SMC is tidally stripped through repeated encounters with the LMC (as illustrated in Extended Data Figs. \ref{fig:clouds} and \ref{fig:temp}b) that occur over a period of 5.7 Gyrs. Because the model includes more massive and more extended disks for the Clouds than previous studies\cite{Besla2012}, these repeated orbits of the SMC around the LMC also result in gas extraction from the LMC\cite{Pardy2018} by dwarf-dwarf galaxy interaction. However, this process acting on both Clouds only contributes 10--20\% of the total Stream mass.

During the early period before the LMC-SMC pair fell into the MW, a Magellanic Corona of gas with $T\sim5\times10^5\unit{K}$ and $M\sim3\times10^9\msol$ surrounded the Magellanic System and extended out to the LMC's virial radius of 100 kpc.
The Corona removes cold gas from the outer disk of the SMC and heats it up by compression, as illustrated in Extended Data Fig. \ref{fig:temp}d. Later the Corona provides an additional source of ionized gas that contributes to the total mass in the Stream. The Corona is therefore a source of pressure, heating, and mass.

\begin{figure}
	\centering
    \includegraphics[width=12.0cm]{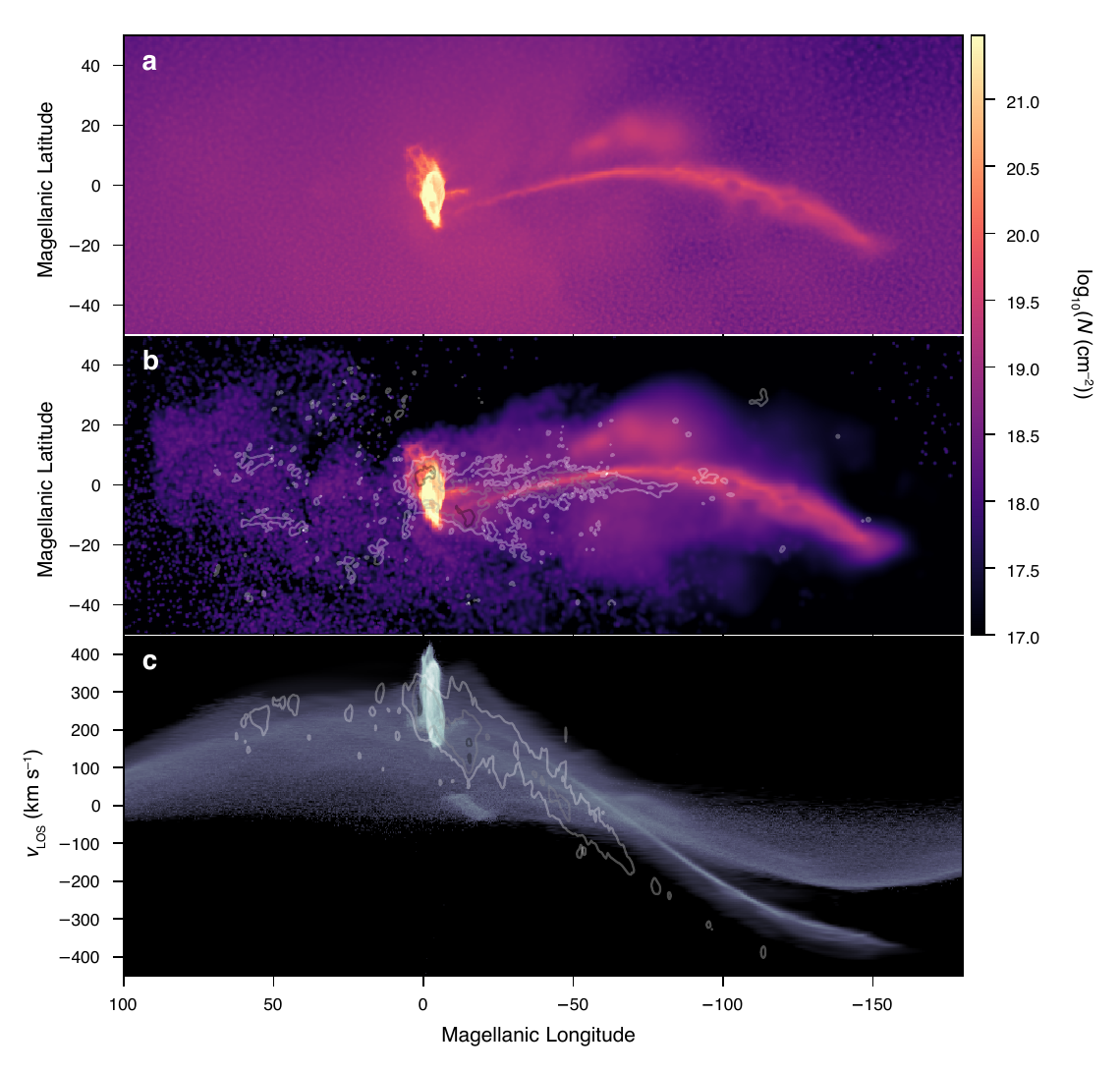}
	\caption{{\bf Gas column density and velocity in Magellanic Coordinates}.
	    {\bf a}, The gas column density of the simulated Stream composed of the Magellanic Corona gas and cold disk gas stripped from the Clouds displayed in Magellanic coordinates. {\bf b}, Column density only of the simulated cold gas Stream as compared to H\,I data\cite{Nidever2010}, with black, gray, white contours corresponding to the observed density of $10^{19}$, $10^{20}$, and $10^{21}$ cm$^{-2}$ respectively. {\bf c}, The line of sight velocity of the total Stream gas as a function of Magellanic longitude, with contours labeled as in {\bf b} and lightness showing relative density.
	}
	\label{fig:magellaniccoords}
\end{figure}

Once the Clouds fell into the MW and the MW hot corona, the Stream was amplified by the MW potential until it extended over 200 degrees in the sky, with both leading and trailing components. Fig. \ref{fig:zea} shows the Stream displayed at the present time in zenithal equal area (ZEA) projection in the numerical experiment (Fig. \ref{fig:zea}b) as compared to the observed Stream (Fig. \ref{fig:zea}a)\cite{McClure-Griffiths2009,Nidever2010}. The MW hot corona included in this model has a total mass of $\sim2\times10^9\msol$ and does not rotate (see Methods). The presence of the hot MW gas and the Magellanic Corona have a large effect on the kinematics of the Stream.
To illustrate this, Fig. \ref{fig:zea}b displays a comparison of line-of-sight velocities of the Stream with the H\,I velocity gradient observed\cite{Nidever2010} in the case when both the Magellanic and MW Coronae are included. The model shows a kinematic gradient from negative to positive velocities along the Stream ($v_\mathrm{LOS}$ from $-$350 to 400 km\,s$^{-1}$), in good agreement with the observed data (Fig. \ref{fig:zea}a). Whereas previous models found the gas to be moving $\sim$100 km\,s$^{-1}$ faster than observations in the LA\cite{Pardy2018} and slower in the Stream\cite{Besla2012}, the inclusion of coronal gas decelerates the LA to better match the observed velocity gradient. However the cold gas column density in this region is smoother than in observations, which indicate the LA is clumpy and fragmented\cite{Bruns2005, Nidever2010} (see Fig. \ref{fig:magellaniccoords}b).

\begin{figure}
	\centering
    \includegraphics[width=13.6cm]{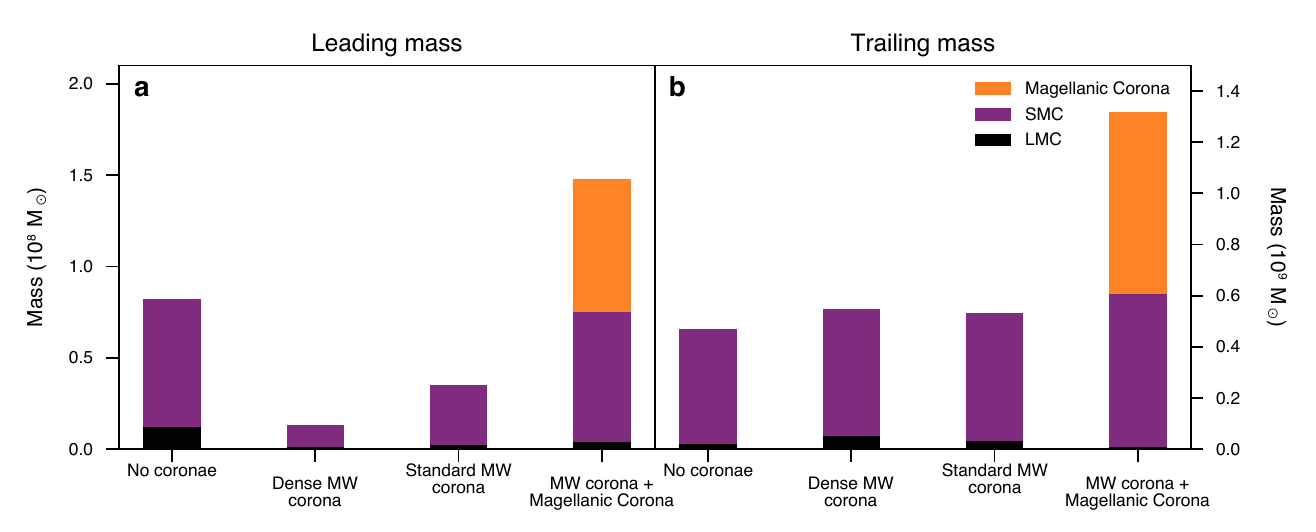}
	\caption{{\bf Stream mass budget}.
		{\bf a}, {\bf b} Origin of the mass in the Leading Arm ({\bf a}) and the Stream ({\bf b}) at present day. Each column represents a model of the formation of the Stream: the fiducial dwarf-dwarf galaxy interaction model (first on the left)\cite{Besla2012,Pardy2018}; a dwarf-dwarf galaxy interaction model with the inclusion of a high-density MW gas halo with total mass $5\times10^{9}$ M$_\odot$ that shows that the Leading Arm does not survive (second left column; see recent work\cite{Tepper-Garcia2019}); a dwarf-dwarf galaxy interaction model with the inclusion of a lower-density MW gas halo (total mass $\sim 2\times10^9$ M$_\odot$) still consistent with current estimates\cite{Bregman2018} (second to the right; see Extended Data Figure \ref{fig:dens}); the model reported here of a dwarf-dwarf galaxy interaction including the lower-density MW gas halo in addition to the Magellanic Corona (right column). The inclusion of the Magellanic Corona shows that this gas contributes greatly to the total mass of the Stream: increasing it to values consistent with observations $\sim1.3\times10^9 \;\mathrm{M_\odot}$.
	}
	\label{fig:bars}
\end{figure}

In our model, both the LMC and the SMC contribute to the formation of the Stream. 
Most of the gas is pulled from the SMC, but there is also a tenuous filamentary contribution from the LMC, 
produced by tidal interactions with the MW and ram-pressure stripping in the MW hot corona.
When the Magellanic System first falls into the MW, the Magellanic Corona is extended. 
Under the influence of the MW gravitational potential, $\sim$22\% of the Magellanic Corona's 
initial mass becomes unbound from the LMC and incorporated into the Stream. Thus by mixing 
with the underlying MW hot gas, the Magellanic Corona contributes to the large ionized mass of 
the Stream. Fig. \ref{fig:bars} shows that the Magellanic Corona contributes
$\sim$50\% of the mass in the Leading Arm and more than 50\% of the total ionized mass in the Stream. The other $\sim$50\% of the mass (in both the Leading Arm and the Stream) is composed of gas extracted earlier from the SMC by its mutual interaction with the LMC with some gas heated by the Magellanic Corona before infall. This additional source of ionized gas has not been accounted for in previous theoretical work and reconciles the Stream's mass budget.

Another outcome of the model concerns the survivability of the Stream and its Leading Arm in the presence of a MW hot corona. 
H\,I studies\cite{Putman1998, Bruns2005} show that the Leading Arm is fragmented,
as expected by simulations of its passage through the Galactic halo\cite{Heitsch2009}, but yet it still survives. However, recent hydrodynamic simulations have challenged the overall survivability of the Leading Arm when the MW hot corona is included\cite{Tepper-Garcia2019}.
The numerical experiment reported here shows that the LA survives if the hot MW halo has a density $n\sim1.7\times10^{-5}\unit{cm^{-3}}$ at a distance of 50 kpc from Galactic center (see Extended Data Fig. \ref{fig:dens}). While the MW corona regulates the formation and morphology of the LA, the inclusion of the Magellanic Corona affects its spatial extent (see Extended Data Fig \ref{fig:mwhaloeffect}). The warm gas surrounding the Clouds provides a shield around the stripped gas to allow the LA gas to penetrate further into the MW hot corona.
Even if the LA turns out to have a non-Magellanic origin, as recently suggested\cite{Tepper-Garcia2019}, the inclusion of the Magellanic Corona still provides the bulk of the mass of the Trailing Stream, including its ionized component.

The inclusion of the Magellanic Corona is further supported by a recent estimate of the ambient gas density near the LA\cite{Nidever2019}. Following the discovery of stars formed in-situ in the LA\cite{Price-Whelan2019}, a recent study\cite{Nidever2019} reports that the density of coronal gas required to separate these young stars from their proposed gaseous nursery (the region known as LA II) is an order of magnitude higher than existing measurements of the MW coronal density\cite{Bregman2018}. This discrepancy can be resolved by taking the Magellanic Corona into account, because the Magellanic Corona can add to the MW corona to yield the high total density needed to ram-pressure-strip the LA II region away from the nascent stars.

An additional consequence of this model is a possible explanation for the lack of a stellar component of the Stream.
In tidal models, stars (in addition to gas) should be stripped from both Clouds due to the gravitational interactions they experience before falling in to the MW.
Such a stellar stream awaits discovery, even though sensitive searches have been conducted. However in our model, the Stream is mostly formed by the warm Magellanic Corona, therefore its stellar counterpart is negligible. Some stars were tidally stripped from the SMC when the Clouds were far from the MW, but they are either phase-mixed with the MW stellar halo or extended into a thin and low-density filament of 30 mag arcsec$^{-2}$, which is too faint to detect with current telescopes and instrumentation.

The Magellanic Corona will be unambiguously observable via absorption in highly-ionized states of carbon and silicon (C\,IV and Si\,IV) in the ultraviolet spectra of background quasars lying near the LMC on the sky. The high-ion column densities in the Corona should decrease with increasing angular separation (impact parameter) from the LMC. In contrast to the ``down-the-barrel'' studies of stars in the LMC\cite{Wakker1998,Lehner2009} which pass through the interstellar medium of the LMC and may probe outflows close to the LMC disk, background-quasar sightlines offer the chance for unambiguous detections of the Corona, because they are uncontaminated by the LMC's interstellar material.

\newpage
\renewcommand{\refname}{Main References}


\newpage
{\Large List of Figures}
\begin{enumerate}
    \item \nameref{fig:zea}
    \item \nameref{fig:magellaniccoords}
    \item \nameref{fig:bars}
\end{enumerate}

\newpage

\section*{Methods}


This work employs the GIZMO Hydrodynamic N-body code\cite{Hopkins2015}. GIZMO includes hydrodynamics schemes that can follow large bulk velocities and large dynamic ranges in density, making it an appropriate tool to model the hydrodynamic evolution of gas disks in isolation and when subjected to gravitational interactions. The Lagrangian
meshless finite-mass method implemented in the code allows the tracking of fluid elements while capturing in detail the Kelvin-Helmholtz instabilities and shocks when the resolution is properly increased\cite{Hopkins2015}.
The simulations also used the adaptive gravitational softening lengths for gas particles available in GIZMO. The softening lengths are determined by the hydrodynamic smoothing lengths to ensure consistency between the gravitational and the hydrodynamic calculations. These smoothing lengths are calculated using the 32 nearest neighbors for each particle. For the dark matter (DM) component, the softening length adopted was 290 pc, and for the stellar component 100 pc was used. The simulations also implemented radiative heating and cooling\cite{Katz1996,Hopkins2018} and star formation and feedback\cite{Springel2003}.

\subsubsection*{Initial set up and simulations.}
We created a set of N-body and hydrodynamic simulations of gaseous and stellar exponential disks embedded in a live NFW (Navarro–Frenk–White) dark matter halo of Magellanic-sized galaxies\cite{Springel2005}.
The LMC progenitor galaxy has a total dark matter halo mass of $17.75\times10^{10}\msol$ ($1.8\times10^5 \msol$ per particle), a stellar mass of $2.5\times10^{9}\msol$ ($4.2\times10^3\msol$ per particle), and a disk gas mass of $2.2\times10^{9}\msol$ ($4.4\times10^3\msol$ per particle). Similarly, the SMC progenitor assumes an initial total dark halo of $2.1\times10^{10}\msol$ ($1.9\times10^5\msol$ per particle), a stellar component of $3\times10^{8}\msol$ ($4.2\times10^3\msol$ per particle), and a gaseous disk of $1.6\times10^{9}\msol$ ($4.4\times10^3\msol$ per particle). This gives $\sim2.6\times10^6$ total particles for the Magellanic Clouds combined.
For the MW, a static Hernquist potential\cite{Hernquist1990} has been assumed with a total mass of $10^{12}\msol$ and a scale length of 29 kpc. A live MW stellar disk and bulge have also been included with masses of $4.8\times10^{10}\msol$ and $8\times10^{9}\msol$ respectively following other recent simulations\cite{D'Onghia2019}. The disk has only been included in the full model with both coronae.

The LMC stellar disk has a scale length of 1.8 kpc while the initial gas disk is extended with a scale length of 4.8 kpc, in agreement with isolated gaseous dwarf irregular galaxies of comparable mass\cite{deBlok1997}.  
Similarly, the scale length of the SMC stellar disk is initially set to 1.1 kpc and the extended gaseous disk has a scale length of 3 kpc. The outer part of the LMC disk is truncated to 25 kpc. Runs performed with the LMC outer disk truncated to various radii produce comparable results. However, for the case reported in this study, the 
filamentary structure of the Trailing Arm from gas tidally removed from the LMC is present but more tenuous and less pronounced as compared to previous work where the LMC disk was not truncated\cite{Pardy2018}.

The Magellanic Corona is set up as a halo of warm gas surrounding the LMC, with a mass of $\sim 3\times10^{9}\msol$ ($\sim$1.5\% of the LMC total mass) and extends throughout the virial radius of the LMC ($\sim$100 kpc). Even though the LMC is a satellite galaxy, it is still massive enough (with total mass $>10^{11}\msol$\cite{Erkal2018,Erkal2019}) to carry a group of dwarfs that includes the SMC, Carina and Fornax\cite{Pardy2019} and several additional ultra-faint dwarfs\cite{Kallivayalil2018}. Hence its hot corona should be at least $10^{9}\msol$ in mass\cite{Hafen2019,Jethwa2016}. A less massive LMC ($\sim 5\times10^{10}\msol$ as inferred from the rotation curve within 8 kpc from the center\cite{Kallivayalil2013}) would not harbor a warm corona and such an LMC would not be massive enough to carry the bright dwarfs as the observations suggest. Cosmological simulations confirm these estimates\cite{Shen2014,Angles-Alcazar2017,Jahn2019}, and dwarf galaxies in the field have been shown to have circumgalactic gas extending out to a significant fraction of their virial radii\cite{Bordoloi2014,Johnson2017,Sokolowska2016}.
Furthermore, a conservative observational estimate of the MW suggests that the circumgalactic gas is at least $\sim$1\% of the total Galactic mass. The observed mass in baryons (stars and the interstellar medium) constitutes $\sim$10\% of the total mass, and it is proposed that the other half of the baryons be found in the hot corona\cite{Fukugita2006}. In addition, absorption line studies show that the mass of the circumgalactic gas inside the virial radius is similar to the stellar mass\cite{Bregman2018,Lehner2007}. Therefore, the total mass of the Magellanic Corona adopted in this work (1.5\% of the LMC mass) should be considered a lower limit.

In this model, the gas properties of the Magellanic Corona surrounding the LMC are extracted from the Auriga simulations\cite{Grand2017}, a set of cosmological simulations of MW-type galaxies that contain LMC-sized satellites. The LMC analogs identified in the Auriga have proper motions similar to the \emph{Hubble Space Telescope} (\emph{HST}) data reported for the Clouds and do have an associated warm gas corona\cite{Pardy2019}, whose properties (temperature of $\sim5\times10^{5}\unit{K}$, density and radial profile) are used as initial conditions for this numerical experiment. 
The density profile (the red dashed line reported in Extended Data Fig. \ref{fig:dens}) decreases at larger radii with a radial profile similar to recent results\cite{Salem2015,Bregman2018,Miller2013} for the MW.
The LMC gas Corona is made up of particles with masses of $4.4\times10^3\msol$. Velocities are assigned to gas particles according to a Maxwell-Boltzmann distribution (as in the isothermal sphere) with $f(v) \propto e^{-\frac{mv^2}{kT}}$ where $m$ is the mean mass per particle, $k$ is Boltzmann's constant, and for $T$ half the virial temperature was assumed.

We note that at $T\sim5\times10^{5}\unit{K}$, the Magellanic Corona is above the peak range of the cooling curve. Although the gaseous coronae in our models are relatively stable due to the inclusion of radiative heating and cooling, star formation and feedback, there may be additional physical processes included in cosmological simulations\cite{Hafen2019,Grand2017}, such as AGN feedback, photoionization heating, and cosmic ray heating, that affect the stability and temperature of the circumgalactic gas\cite{Bustard2018,Bustard2019,Gronke2019}.

In addition, a gas corona was set up around the Milky Way assuming an isothermal sphere of gas at $T=1.6\times10^{6} \unit{K}$ (the Galactic virial temperature) using the DICE code\cite{dice}. The MW gas corona does not rotate in this model and we find that the infall of the Magellanic System does not affect the large-scale rotation of the coronal gas. As shown in previous work\cite{Tepper-Garcia2019}, the rotation of the MW hot corona can have effects on the morphology and structure of the Stream, however for this study we are investigating the macroscopic properties of the Stream which should not be affected by the MW corona's rotation.
The hot corona has a total mass of $\sim 2\times10^9\msol$ made up of particles with masses of $4.5\times10^3\msol$. It was allowed to equilibrate in isolation (with the static MW DM potential) for $\sim1$ Gyr before the MCs fell in. The gas density profile assumed for the final run follows the distribution reported in previous studies\cite{Salem2015,Bregman2018,Faerman2019} and is displayed in Extended Data Fig. \ref{fig:dens} (solid red line). The Magellanic Corona and MW hot gas corona constitute an additional $2\times10^6$ particles in the simulation.

\subsubsection*{Magellanic Cloud Orbital Parameters.}
A parameter study of the orbital configurations of the Clouds was carried out.
Consistent with the findings of previous works\cite{Diaz2012,Besla2010,Besla2012,Pardy2018, D'Onghia2009,D'Onghia2010}, the orbits for the LMC and SMC were set such that the Clouds experience three mutual gravitational encounters before falling into the MW potential. Note that the orbital configuration parameters were set to reproduce the bifurcation of the Stream and the H\,I component which is only 10-20\% of its total mass. In the model shown here the Magellanic Corona is the dominant source of the total Stream mass. This result is independent on the number of encounters between the Clouds and their structural parameters.
The LMC orbit is obtained first by solving the differential equation of motion assuming a mass of $2\times 10^{11}$ M$_{\odot}$ for the LMC before the infall and a MW mass of $\sim10^{12}$ M$_{\odot}$. By imposing the current observed velocities and positions for the LMC as inferred by \emph{HST} data, differential equations of motion are used to determine the position and velocities of the LMC at earlier times. Following previous studies\cite{Pardy2018} the SMC is initially placed $65\kpc$ away from the LMC on a Keplerian orbit with eccentricity $e=0.65$, and minimum separation of $25\kpc$ from the LMC. The orbital history of the Clouds and their mutual interactions away from the Milky Way are illustrated in Extended Data Fig. \ref{fig:clouds}. 

After the Clouds have had three close encounters, over a time period of 5.7 Gyrs, the LMC and SMC are placed 220 kpc away from the center of the MW on a first pericentric passage around the Galaxy.
The LMC-SMC system is rotated by $180^\circ$ around the $z$-axis, then $100^\circ$ around the $y$-axis, then $-50^\circ$ around the $x$-axis. Then the LMC's center of mass is placed at $(x,y,z) = (-22,217,32)$ kpc (where the MW hot corona and DM potential are centered at the origin) with a velocity of $(v_x, v_y, v_z) = (18.6, -88.6, -109)\unit{km/s}$. The SMC's position and velocity were unchanged relative to the LMC for the first 5.7 Gyrs in isolation.
Once the Clouds fall into the MW, they reach their present day positions after 1.3 Gyrs with velocities consistent with current observations\cite{Kallivayalil2006,Kallivayalil2013}. The Stream 
at present day is displayed in ZEA coordinates in Extended Data Fig. \ref{fig:mwhaloeffect}. A fiducial model where the Stream is formed by the mutual interaction between the Clouds without the inclusion of the warm and hot corona was run first (Extended Data Fig. \ref{fig:mwhaloeffect}a)\cite{Pardy2018}. Subsequently, the same model assumed for the Clouds is carried out with the inclusion of a high-density (Extended Data Fig. \ref{fig:mwhaloeffect}b) or low-density MW hot gas corona (Extended Data Fig. \ref{fig:mwhaloeffect}c,d). This experiment allowed us to determine that the Leading Arm survives in this model if the MW gas corona has a density of $n\sim1.7\times10^{-5}\unit{cm^{-3}}$ at a distance of 50 kpc, 
in agreement with the observational estimates\cite{Salem2015} and previous studies\cite{Tepper-Garcia2019}. The final run included the model of the Clouds with the inclusion of both the Magellanic Corona and the MW hot halo (Extended Data Fig. \ref{fig:mwhaloeffect}d).  

\subsubsection*{Analysis.}

A particle tracer that allows us to follow each gas particle with its temperature and density was employed to compute the mass of the Magellanic Stream.
In these numerical experiments the Stream consists of gas particles stripped from the Clouds that are no longer bound to the main body of their host galaxy. The gravitational potential and its kinetic energy were calculated for each gas particle. Any particle that has a larger kinetic than potential energy was considered unbound.  We then projected the locations of gas particles stripped from the Clouds into Magellanic Coordinates and summed up the masses.
Based on the locations of the gas particles, they were included either in the Leading or the Trailing Stream. The pygad\cite{pygad} library was adopted to perform density and temperature calculations and to deposit the particles onto a mesh for visualization.
The model does not include the ionization corrections to convert the hydrogen gas into the ionized fraction. The cold gas stripped from the Clouds is assumed to trace the HI component, whereas the warm coronal gas is assumed to trace the ionized mass.


\newpage
\renewcommand{\refname}{Methods References}

\newpage

{\small 
\noindent
\textbf{Data Availability} The GIZMO code used in this work is available from https://bitbucket.org/phopkins/gizmo-public/. The PyGad code used in this work is available from https://bitbucket.org/broett/pygad/. The simulation data that support the findings are available from GitHub, https://github.com/DOnghiaGroup/lucchini-2020-sim/.

\noindent
\textbf{Acknowledgments} E.D. acknowledges the hospitality of the Center for Computational Astrophysics at the Flatiron Institute during the completion of this work. 

\noindent
\textbf{Author Contributions} S.L., E.D., and A.J.F. conceived and developed the numerical experiments. J.B., C.B., and E.Z. contributed to the discussion of the physical processes and all authors assisted with editing of the manuscript. S.L. created the figures with input from C.B.

\noindent
\textbf{Competing Interests} The authors declare no competing interests.

\noindent
\textbf{Additional Information} Correspondence and requests for materials should be addressed to E.D. (email: edonghia@astro.wisc.edu). Reprints and permissions information is available at www.nature.com/reprints.
}

\renewcommand{\figurename}{Extended Data Figure}
\setcounter{figure}{0}

\begin{figure}[ph]
	\centering
    \includegraphics[width=12.0cm]{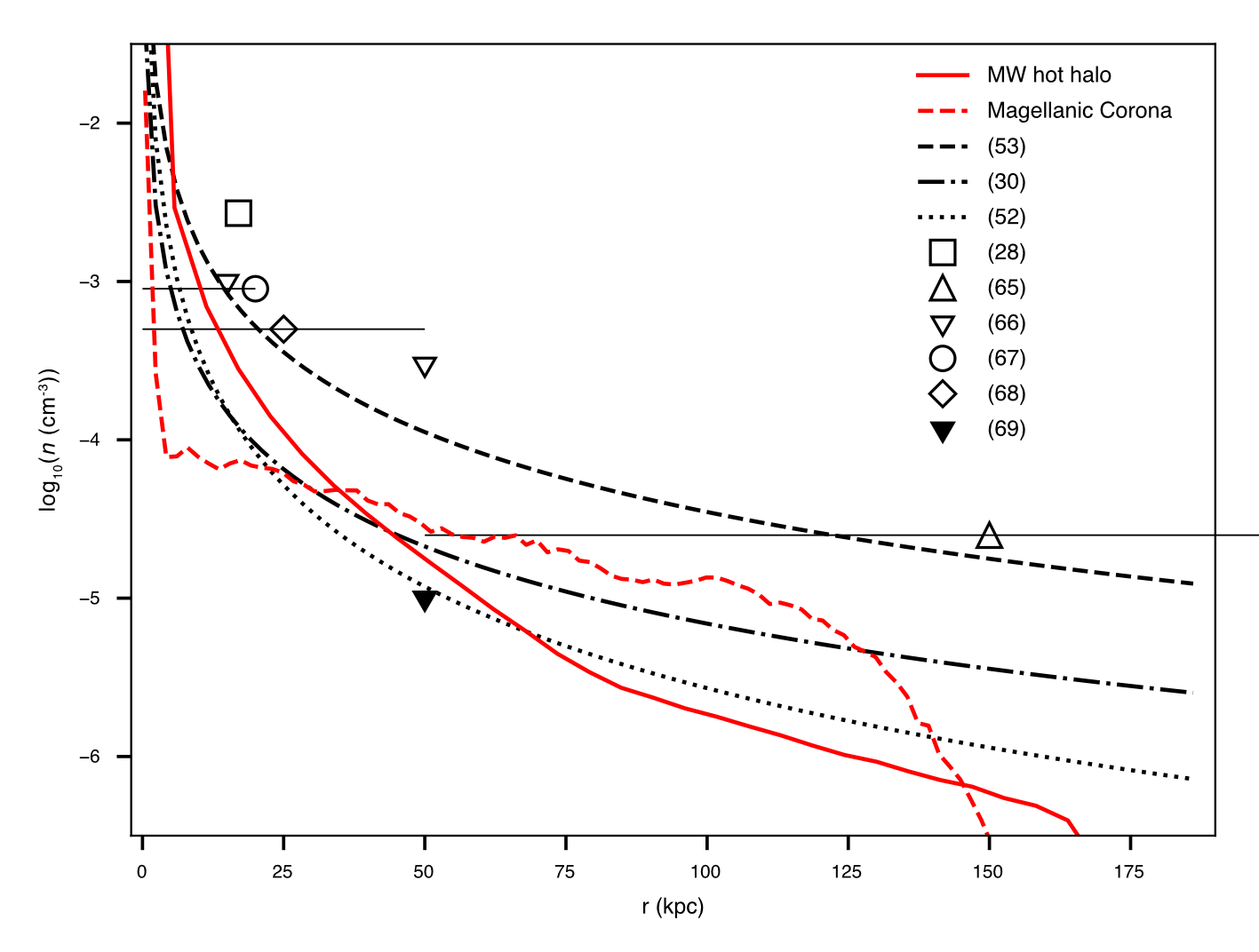}
	\caption{{\bf Radial gas density profile of the Magellanic Corona and MW hot corona}.
		The number density of gas in the models of the Magellanic Corona (marked by the dashed red line) and the MW hot corona (the solid red line) are shown as a function of radius (from the center of the LMC and MW respectively). Estimates of the MW hot coronal density from observations are shown in black. The dotted and dot-dashed lines correspond to the functional form fit to data\cite{Salem2015,Bregman2018,Miller2013}.
		The data points are numbered with their corresponding references, and are the same as those included in previous studies\cite{Tepper-Garcia2019}. Downward (upward) pointing triangles indicate upper (lower) limits.
	}
	\label{fig:dens}
\end{figure}

\begin{figure}[ph]
	\centering
    \includegraphics[width=\textwidth]{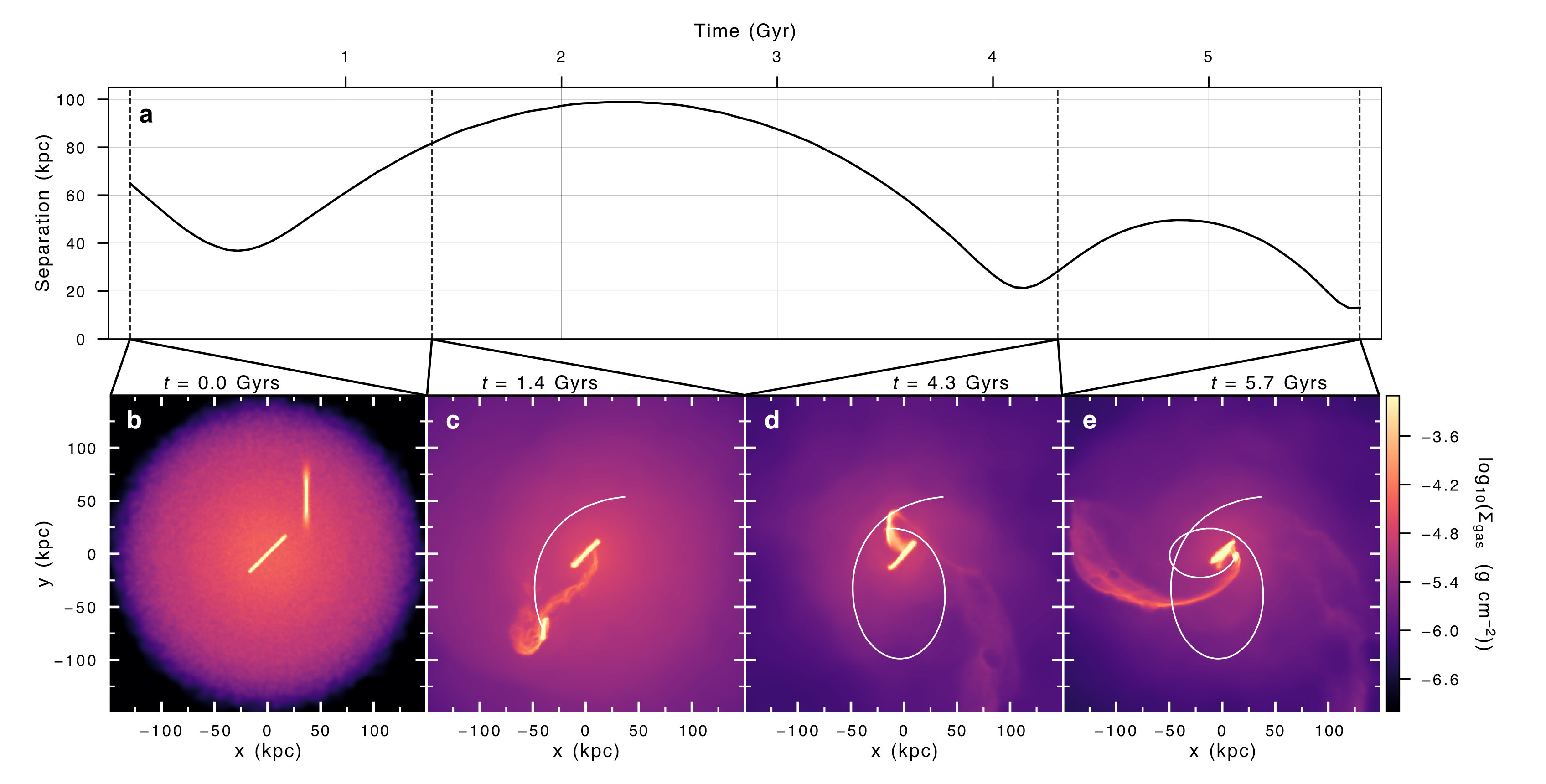}
	\caption{{\bf Orbital histories of the Large and Small Magellanic Clouds}.
		{\bf a}, Time evolution of the distance between the center of mass of the LMC and the center of mass of the SMC. The clouds interact gravitationally for a period of 5.7 Gyrs (three close encounters) before falling into the MW potential.	{\bf b-e}, Gas column density at various times during the Clouds' mutual interactions (at the initial time, after 1.4, 4.3, and 5.7 Gyrs; marked on the top plot with dotted vertical lines). Displayed is the gas tidally removed from the LMC and SMC in addition to the Magellanic Coronal gas.
	}
	\label{fig:clouds}
\end{figure}

\begin{figure}[ph]
	\centering
    \includegraphics[width=13.8cm]{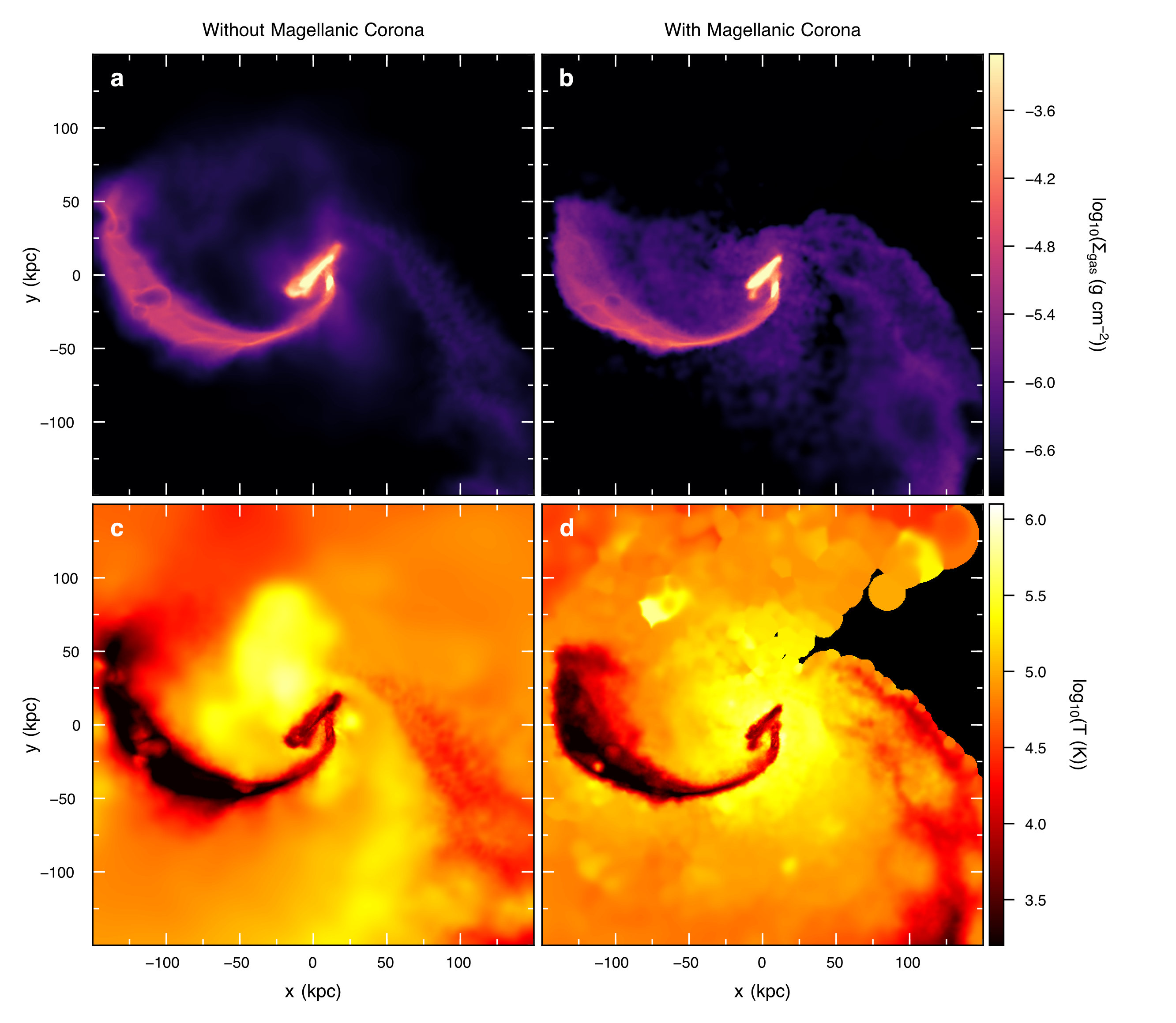}
    \caption{{\bf The effect of the Magellanic Corona on stripped gas temperature}.
		The gas removed from the Magellanic Clouds after $\sim 5.7$ Gyrs of mutual interactions (before infall into the MW potential) is shown in Cartesian coordinates projected along the z-axis onto the x-y plane. The LMC and SMC are at the center of each panel.
		{\bf a, b}, The gas mass surface density of the gas originating the the disks of the Clouds.
		{\bf c, d}, The gas temperature averaged along the projection axis.
		Shown both for models run with ({\bf b, d}) and without ({\bf a, c}) the Magellanic Corona included.
	}
	\label{fig:temp}
\end{figure}

\begin{figure}[ph]
	\centering
    \includegraphics[width=12.0cm]{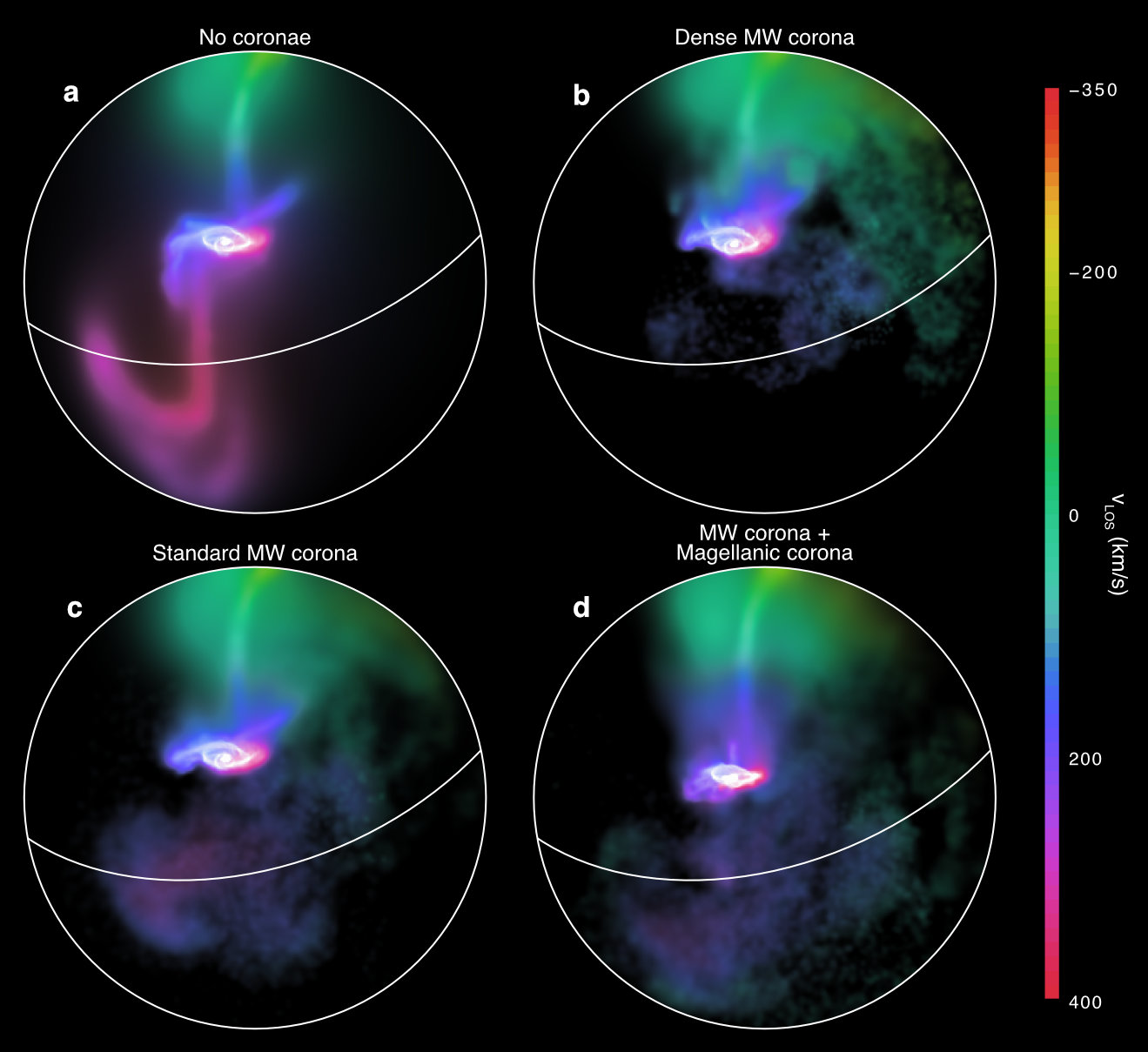}
	\caption{{\bf The effect of the warm and hot gas on the formation of the Leading Arm}.
	    Column density (brightness) and line-of-sight velocity (color) for four different models for the formation of the Magellanic Stream. These four models are the same as those in Fig. \ref{fig:bars} of the main article. 
	    In all four plots only the gas originating in the gaseous disks of the Magellanic Clouds is displayed.
	    {\bf a}, Fiducial model, without the MW corona or Magellanic Corona (tidal forces only).
	    {\bf b}, A MW coronal mass of $5\times10^{9}\msol$ is included but the Magellanic Corona is not present. The Leading Arm does not survive, in agreement with previous studies\cite{Tepper-Garcia2019}. 
	    {\bf c}, Same as {\bf b} except with the total mass of the MW hot corona lowered to $2\times10^9\msol$ (see Extended Data Fig. \ref{fig:dens}), allowing the Leading Arm to survive.  {\bf d}, Same as {\bf c} except with the addition of the Magellanic Corona. This model provides the best match to observations.
	}
	\label{fig:mwhaloeffect}
\end{figure}


\begin{thebibliography}{10}

\def\mnras{Mon. Not. R. Astron. Soc.}
\def\apj{Astrophys. J.}
\def\apjl{Astrophys. J. Lett.}
\def\apjs{Astrophys. J. Supp.}
\def\aap{Astron. Astrophys.}
\def\araa{Annu. Rev. Astron. Astrophys.}


\bibitem{Mathewson1974} Mathewson, D.~S., Cleary, M.~N., \& Murray, J.~D. The Magellanic Stream. \emph{\apj}\ \textbf{190}, 291-296 (1974).

\bibitem{Nidever2008} Nidever, D.~L., Majewski, S.~R., \& Butler Burton, W. The Origin of the Magellanic Stream and Its Leading Arm. \emph{\apj}\ \textbf{679}, 432-459 (2008).

\bibitem{Kallivayalil2013} Kallivayalil, N., van der Marel, R.~P., Besla, G., Anderson, J., \& Alcock, C. Third-epoch Magellanic Cloud Proper Motions. I. Hubble Space Telescope/WFC3 Data and Orbit Implications. \emph{\apj}\ \textbf{764}, 161 (2013).

\bibitem{Fox2013} Fox, A.~J., et al. The COS/UVES Absorption Survey of the Magellanic Stream. I. One-tenth Solar Abundances along the Body of the Stream. \emph{\apj}\ \textbf{772}, 110 (2013).

\bibitem{Richter2013} Richter, P., et al. The COS/UVES Absorption Survey of the Magellanic Stream. II. Evidence for a Complex Enrichment History of the Stream from the Fairall 9 Sightline. \emph{\apj}\ \textbf{772}, 111 (2013).

\bibitem{Fox2014} Fox, A.~J., et al. The COS/UVES Absorption Survey of the Magellanic Stream. III. Ionization, Total Mass, and Inflow Rate onto the Milky Way. \emph{\apj}\ \textbf{787}, 147 (2014).

\bibitem{Barger2017} Barger, K.~A., et al. Revealing the Ionization Properties of the Magellanic Stream Using Optical Emission. \emph{\apj}\ \textbf{851}, 110 (2017).

\bibitem{Besla2012} Besla, G., et al. The role of dwarf galaxy interactions in shaping the Magellanic System and implications for Magellanic Irregulars. \emph{\mnras}\ \textbf{421}, 2109-2138 (2012).

\bibitem{Pardy2018} Pardy, S.~A., D'Onghia, E., \& Fox, A.~J. Models of Tidally Induced Gas Filaments in the Magellanic Stream. \emph{\apj}\ \textbf{857}, 101 (2018).

\bibitem{Hammer2015} Hammer, F., Yang, Y.~B., Flores, H., Puech, M., \& Fouquet, S. The Magellanic Stream System. I. Ram-Pressure Tails and the Relics of the Collision Between the Magellanic Clouds. \emph{\apj}\ \textbf{813}, 110 (2015).

\bibitem{Wang2019} Wang, J., et al. Towards a complete understanding of the Magellanic Stream Formation. \emph{\mnras}\ \textbf{486}, 5907-5916 (2019).

\bibitem{D'Onghia2016} D'Onghia, E., \& Fox, A.~J. The Magellanic Stream: Circumnavigating the Galaxy. \emph{\araa}\ \textbf{54}, 363-400 (2016).

\bibitem{Penarrubia2016} Pe{\~n}arrubia, J., G{\'o}mez, F.~A., Besla, G., Erkal, D., \& Ma, Y.-Z. A timing constraint on the (total) mass of the Large Magellanic Cloud. \emph{\mnras}\ \textbf{456}, L54-L58 (2016).

\bibitem{Wakker1998} Wakker, B., Howk, J.~C., Chu, Y.-H., Bomans, D., \& Points, S.~D. Coronal C$^{+3}$ in the Large Magellanic Cloud: Evidence for a Hot Halo. \emph{\apjl}\ \textbf{499}, L87-L91 (1998).

\bibitem{Lehner2009} Lehner, N., Staveley-Smith, L., \& Howk, J.~C. Properties and Origin of the High-Velocity Gas Toward the Large Magellanic Cloud. \emph{\apj}\ \textbf{702}, 940-954 (2009).


\bibitem{Bruns2005} Br{\"u}ns, C., et al. The Parkes H I Survey of the Magellanic System. \emph{\aap}\ \textbf{432}, 45-67 (2005).

\bibitem{Nidever2010} Nidever, D.~L., Majewski, S.~R., Butler Burton, W., \& Nigra, L. The 200$^\circ$ Long Magellanic Stream System. \emph{\apj}\ \textbf{723}, 1618-1631 (2010).

\bibitem{Besla2007} Besla, G., et al. Are the Magellanic Clouds on Their First Passage about the Milky Way? \emph{\apj}\ \textbf{668}, 949-967 (2007).

\bibitem{Bechtol2015} Bechtol, K., et al. Eight New Milky Way Companions Discovered in First-year Dark Energy Survey Data. \emph{\apj}\ \textbf{807}, 50 (2015).

\bibitem{D'Onghia2008} D'Onghia, E., \& Lake, G. Small Dwarf Galaxies within Larger Dwarfs: Why Some Are Luminous while Most Go Dark. \emph{\apjl}\ \textbf{686}, L61 (2008).

\bibitem{Nichols2011} Nichols, M., Colless, J., Colless, M., \& Bland-Hawthorn, J. Accretion of the Magellanic System onto the Galaxy. \emph{\apj}\ \textbf{742}, 110 (2011).

\bibitem{Pardy2019} Pardy, S.~A., et al. Satellites of Satellites: The Case for Carina and Fornax. \emph{\mnras}\ 2789 (2019).

\bibitem{Hafen2019} Hafen, Z., et al. The origins of the circumgalactic medium in the FIRE simulations. \emph{\mnras}\ \textbf{488}, 1248-1272 (2019).

\bibitem{McClure-Griffiths2009} McClure-Griffiths, N.~M., et al. Gass: The Parkes Galactic All-Sky Survey. I. Survey Description, Goals, and Initial Data Release. \emph{\apjs}\ \textbf{181}, 398-412 (2009).

\bibitem{Putman1998} Putman, M. E. et al. Tidal disruption of the Magellanic Clouds by the Milky Way. \emph{Nature} \textbf{394}, 752 (1998).

\bibitem{Heitsch2009} Heitsch, F., \& Putman, M.~E. The Fate of High-Velocity Clouds: Warm or Cold Cosmic Rain?. \emph{\apj}\ \textbf{698}, 1485-1496 (2009).

\bibitem{Tepper-Garcia2019} Tepper-Garc{\'\i}a, T., Bland-Hawthorn, J., Pawlowski, M.~S., \& Fritz, T.~K. The Magellanic System: the puzzle of the leading gas stream. \emph{\mnras}\ \textbf{488}, 918-938 (2019).

\bibitem{Nidever2019} Nidever, D.~L., et al. Spectroscopy of the Young Stellar Association Price-Whelan 1: Origin in the Magellanic Leading Arm and Constraints on the Milky Way Hot Halo. \emph{\apj}\ \textbf{887}, 115 (2019).

\bibitem{Price-Whelan2019} Price-Whelan, A.~M., et al. Discovery of a Disrupting Open Cluster Far into the Milky Way Halo: A Recent Star Formation Event in the Leading Arm of the Magellanic Stream?. \emph{\apj}\ \textbf{887}, 19 (2019).

\bibitem{Bregman2018} Bregman, J.~N., et al. The Extended Distribution of Baryons around Galaxies. \emph{\apj}\ \textbf{862}, 3 (2018).


\end{thebibliography}

\begin{thebibliography}{10}

\makeatletter
\addtocounter{\@listctr}{30}
\makeatother

\def\mnras{Mon. Not. R. Astron. Soc.}
\def\apj{Astrophys. J.}
\def\apjs{Astrophys. J. Supp.}
\def\apjl{Astrophys. J. Lett.}
\def\aap{Astron. Astrophys.}

\bibitem{Hopkins2015} Hopkins, P.~F. A new class of accurate, mesh-free hydrodynamic simulation methods. \emph{\mnras}\ \textbf{450}, 53-110 (2015).

\bibitem{Katz1996} Katz, N., Weinberg, D.~H., \& Hernquist, L. Cosmological Simulations with TreeSPH. \emph{\apjs}\ \textbf{105}, 19 (1996).

\bibitem{Hopkins2018} Hopkins, P.~F., et al. FIRE-2 simulations: physics versus numerics in galaxy formation. \emph{\mnras}\ \textbf{480}, 800-863 (2018).

\bibitem{Springel2003} Springel, V., \& Hernquist, L. Cosmological smoothed particle hydrodynamics simulations: a hybrid multiphase model for star formation. \emph{\mnras}\ \textbf{339}, 289-311 (2003).

\bibitem{Springel2005} Springel, V. The cosmological simulation code GADGET-2. \emph{\mnras}\ \textbf{364}, 1105-1134 (2005).

\bibitem{Hernquist1990} Hernquist, L. An Analytical Model for Spherical Galaxies and Bulges. \emph{\apj}\ \textbf{356}, 359 (1990).

\bibitem{D'Onghia2019} D'Onghia, E., \& Aguerri, J.~A.~L. Trojans in the Solar Neighborhood. \emph{arXiv e-prints}\ \textbf arXiv:1907.08484 (2019).

\bibitem{deBlok1997} de Blok, W.~J.~G., \& McGaugh, S.~S. The dark and visible matter content of low surface brightness disc galaxies. \emph{\mnras}\ \textbf{290}, 533-552 (1997).

\bibitem{Erkal2018} Erkal, D., et al. Modelling the Tucana III stream - a close passage with the LMC. \emph{\mnras}\ \textbf{481}, 3148-3159 (2018).

\bibitem{Erkal2019} Erkal, D., et al. The total mass of the Large Magellanic Cloud from its perturbation on the Orphan stream. \emph{\mnras}\ \textbf{487}, 2685-2700 (2019).

\bibitem{Kallivayalil2018} Kallivayalil, N., et al. The Missing Satellites of the Magellanic Clouds? Gaia Proper Motions of the Recently Discovered Ultra-faint Galaxies. \emph{\apj}\ \textbf{867}, 19 (2018).

\bibitem{Jethwa2016} Jethwa, P., Erkal, D., \& Belokurov, V. A Magellanic origin of the DES dwarfs. \emph{\mnras}\ \textbf{461}, 2212-2233 (2016).

\bibitem{Shen2014} Shen, S., Madau, P., Conroy, C., Governato, F., \& Mayer, L. The Baryon Cycle of Dwarf Galaxies: Dark, Bursty, Gas-rich Polluters. \emph{\apj}\ \textbf{792}, 99 (2014).

\bibitem{Angles-Alcazar2017} Angl{\'e}s-Alc{\'a}zar, D., et al. The cosmic baryon cycle and galaxy mass assembly in the FIRE simulations. \emph{\mnras}\ \textbf{470}, 4698-4719 (2017).

\bibitem{Jahn2019} Jahn, E.~D., et al. Dark and luminous satellites of LMC-mass galaxies in the FIRE simulations. \emph{\mnras}\ \textbf{489}, 5348-5364 (2019).

\bibitem{Bordoloi2014} Bordoloi, R., et al. The COS-Dwarfs Survey: The Carbon Reservoir around Sub-L* Galaxies. \emph{\apj}\ \textbf{796}, 136 (2014).

\bibitem{Johnson2017} Johnson, S.~D., Chen, H.-W., Mulchaey, J.~S., Schaye, J., \& Straka, L.~A. The Extent of Chemically Enriched Gas around Star-forming Dwarf Galaxies. \emph{\apjl}\ \textbf{850}, L10 (2017).

\bibitem{Sokolowska2016} Soko{\l}owska, A., Mayer, L., Babul, A., Madau, P., \& Shen, S. Diffuse Coronae in Cosmological Simulations of Milky Way-sized Galaxies. \emph{\apj}\ \textbf{819}, 21 (2016).

\bibitem{Fukugita2006} Fukugita, M., \& Peebles, P.~J.~E. Massive Coronae of Galaxies. \emph{\apj}\ \textbf{639}, 590-599 (2006).

\bibitem{Lehner2007} Lehner, N., \& Howk, J.~C. Highly ionized plasma in the Large Magellanic Cloud: evidence for outflows and a possible galactic wind. \emph{\mnras}\ \textbf{377}, 687-704 (2007).

\bibitem{Grand2017} Grand, R.~J.~J., et al. The Auriga Project: the properties and formation mechanisms of disc galaxies across cosmic time. \emph{\mnras}\ \textbf{467}, 179-207 (2017).

\bibitem{Miller2013} Miller, M.~J., \& Bregman, J.~N. The Structure of the Milky Way's Hot Gas Halo. \emph{\apj}\ \textbf{770}, 118 (2013).

\bibitem{Salem2015} Salem, M., et al. Ram pressure stripping of the Large Magellanic Cloud's disk as a probe of the Milky Way's circumgalactic medium. \emph{\apj}\ \textbf{815}, 77 (2015).

\bibitem{Bustard2018} Bustard, C., Pardy, S.~A., D'Onghia, E., Zweibel, E.~G., \& Gallagher, J.~S. The Fate of Supernova-heated Gas in Star-forming Regions of the LMC: Lessons for Galaxy Formation?. \emph{\apj}\ \textbf{863}, 49 (2018).

\bibitem{Bustard2019} Bustard, C., Zweibel, E.~G., D'Onghia, E., Gallagher, J.~S., \& Farber, R. Cosmic Ray Driven Outflows from the Large Magellanic Cloud: Contributions to the LMC Filament. \emph{arXiv e-prints}\ \textbf arXiv:1911.02021 (2019).

\bibitem{Gronke2019} Gronke, M., \& Oh, S.~P. How cold gas continuously entrains mass and momentum from a hot wind. \emph{\mnras}\ 2995 (2019).

\bibitem{dice} Perret, V., et al. Evolution of the mass, size, and star formation rate in high redshift merging galaxies. MIRAGE - A new sample of simulations with detailed stellar feedback. \emph{\aap}\ \textbf{562}, A1 (2014).

\bibitem{Faerman2019} Faerman, Y., Sternberg, A., McKee, C.~F. Massive Warm/Hot Galaxy Coronae: II -- Isentropic Model. Preprint at https://arxiv.org/abs/1909.09169 (2019).

\bibitem{Besla2010} Besla, G., et al. Simulations of the Magellanic Stream in a First Infall Scenario. \emph{\apjl}\ \textbf{721}, L97-L101 (2010).

\bibitem{D'Onghia2009} D'Onghia, E., Besla, G., Cox, T.~J., \& Hernquist, L. Resonant stripping as the origin of dwarf spheroidal galaxies. {\it Nature} \textbf{460}, 605-607 (2009).

\bibitem{D'Onghia2010} D'Onghia, E., Vogelsberger, M., Faucher-Giguere, C.-A., \& Hernquist, L. Quasi-resonant Theory of Tidal Interactions. \emph{\apj}\ \textbf{725}, 353-368 (2010).

\bibitem{Diaz2012} Diaz, J.~D., \& Bekki, K. The Tidal Origin of the Magellanic Stream and the Possibility of a Stellar Counterpart. \emph{\apj}\ \textbf{750}, 36 (2012).

\bibitem{Kallivayalil2006} Kallivayalil, N., et al. The Proper Motion of the Large Magellanic Cloud Using HST. \emph{\apj}\ \textbf{638}, 772-785 (2006).

\bibitem{pygad} R{\"o}ttgers, B. pygad: Analyzing Gadget Simulations with Python. \emph{Astrophysics Source Code Library}\ ascl:1811.014 (2018).

\bibitem{Blitz2000} Blitz, L., \& Robishaw, T. Gas-Rich Dwarf Spheroidals. \emph{\apj}\ \textbf{541}, 675-687 (2000).

\bibitem{Stanimirovic2002} Stanimirovi{\'c}, S., Dickey, J.~M., Kr{\v{c}}o, M., \& Brooks, A.~M. The Small-Scale Structure of the Magellanic Stream. \emph{\apj}\ \textbf{576}, 773-789 (2002).

\bibitem{Bregman2007} Bregman, J.~N., \& Lloyd-Davies, E.~J. X-Ray Absorption from the Milky Way Halo and the Local Group. \emph{\apj}\ \textbf{669}, 990-1002 (2007).

\bibitem{Anderson2010} Anderson, M.~E., \& Bregman, J.~N. Do Hot Halos Around Galaxies Contain the Missing Baryons?. \emph{\apj}\ \textbf{714}, 320-331 (2010).

\bibitem{Murali2000} Murali, C. The Magellanic Stream and the Density of Coronal Gas in the Galactic Halo. \emph{\apjl}\ \textbf{529}, L81-L84 (2000).

\end{thebibliography}
\end{document}